\documentclass{emulateapj}

\usepackage{natbib}
\usepackage{amsmath}
\usepackage[backref,breaklinks,colorlinks,citecolor=blue]{hyperref}
\usepackage[all]{hypcap}

\newcommand{\about}{$\sim\!\!$~}

\def\lsim{\hbox{\rlap{\raise 0.425ex\hbox{$<$}}\lower 0.65ex\hbox{$\sim$}}}
\def\gsim{\hbox{\rlap{\raise 0.425ex\hbox{$>$}}\lower 0.65ex\hbox{$\sim$}}}

\shorttitle{Progenitor of SN~2014dt}
\shortauthors{Foley et~al.}

\begin{document}

\title{On the Progenitor System of the Type~I\lowercase{ax} Supernova 2014\lowercase{dt} in M61}

\def\illast{1}
\def\illphy{2}
\def\ipac{3}
\def\rut{4}
\def\berk{5}
\def\jpl{6}
\def\cit{7}
\def\az{8}
\def\hf{$\dagger$}

\author{
{Ryan~J.~Foley}\altaffilmark{\illast,\illphy},
{Schuyler~D.~Van~Dyk}\altaffilmark{\ipac},
{Saurabh~W.~Jha}\altaffilmark{\rut},
{Kelsey~I.~Clubb}\altaffilmark{\berk},
{Alexei~V.~Filippenko}\altaffilmark{\berk},
{Jon~C.~Mauerhan}\altaffilmark{\berk},
{Adam~A.~Miller}\altaffilmark{\jpl,\cit,\hf}, and
{Nathan~Smith}\altaffilmark{\az}
}

\altaffiltext{\illast}{
Astronomy Department,
University of Illinois at Urbana-Champaign,
1002 W.\ Green Street,
Urbana, IL 61801, USA
}
\altaffiltext{\illphy}{
Department of Physics,
University of Illinois Urbana-Champaign,
1110 W.\ Green Street,
Urbana, IL 61801, USA
}
\altaffiltext{\ipac}{
IPAC/Caltech,
Mail Code 100-22,
Pasadena, CA 91125, USA
}
\altaffiltext{\rut}{
Department of Physics and Astronomy,
Rutgers, The State University of New Jersey,
136 Frelinghuysen Road,
Piscataway, NJ 08854, USA
}
\altaffiltext{\berk}{
Department of Astronomy,
University of California,
Berkeley, CA 94720-3411, USA
}
\altaffiltext{\jpl}{
Jet Propulsion Laboratory,
California Institute of Technology,
4800 Oak Grove Drive,
MS 169-506,
Pasadena, CA 91109, USA
}
\altaffiltext{\cit}{
California Institute of Technology,
Pasadena, CA 91125, USA
}
\altaffiltext{\az}{
Steward Observatory,
University of Arizona,
Tucson, AZ 85721, USA
}
\altaffiltext{\hf}{
Hubble Fellow
}

\begin{abstract}
  We present pre-explosion and post-explosion {\it Hubble Space
    Telescope} images of the Type Iax supernova (SN~Iax) 2014dt in
  M61.  After astrometrically aligning these images, we do not detect
  any stellar sources at the position of the SN in the pre-explosion
  images to relatively deep limits (3$\sigma$ limits of $M_{\rm F438W}
  > -5.0$~mag and $M_{\rm F814W} > -5.9$~mag).  These limits are
  similar to the luminosity of SN~2012Z's progenitor system ($M_{\rm
    F435W} = -5.43 \pm 0.15$ and $M_{\rm F814W} = -5.24 \pm
  0.16$~mag), the only probable detected progenitor system in
  pre-explosion images of a SN Iax, and indeed, of any white dwarf
  supernova.  SN~2014dt is consistent with having a C/O white-dwarf
  primary/helium-star companion progenitor system, as was suggested
  for SN~2012Z, although perhaps with a slightly smaller or hotter
  donor.  The data are also consistent with SN~2014dt having a
  low-mass red giant or main-sequence star companion.  The data rule
  out main-sequence stars with $M_{\rm init} \gtrsim 16 M_{\sun}$ and
  most evolved stars with $M_{\rm init} \gtrsim 8 M_{\sun}$ as being
  the progenitor of SN~2014dt.  Hot Wolf-Rayet stars are also allowed,
  but the lack of nearby bright sources makes this scenario unlikely.
  Because of its proximity ($D = 12$~Mpc), SN~2014dt is ideal for
  long-term monitoring, where images in \about 2 years may detect the
  companion star or the luminous bound remnant of the progenitor white
  dwarf.
\end{abstract}


%%%%%%%%%%%%%%%%%%%%
%%  Introduction  %%
%%%%%%%%%%%%%%%%%%%%

\defcitealias{McCully14:12z}{M14a}
\defcitealias{Foley13:iax}{F13}

\section{Introduction}\label{s:intro}

Type Iax supernovae (SNe~Iax) are a recently defined class of stellar
explosion \citep[hereafter \citetalias{Foley13:iax}]{Foley13:iax}.
They are, in many ways, observationally similar to SNe~Ia, having
comparable spectra and thus compositions \citep[e.g.,][]{Li03:02cx,
  Branch05, Chornock06, Jha06:02cx, Foley09:08ha, Foley10:08ha}.
However, SNe~Iax are less energetic, with ejecta velocities near
maximum light 20--80\% that of typical SNe~Ia (e.g.,
\citealt{Narayan11}; \citetalias{Foley13:iax}; \citealt{White14}).
SNe~Iax also tend to have lower luminosity than SNe~Ia
\citep[e.g.,][]{McClelland10, Stritzinger14:10ae, Stritzinger14:12z},
further indicating a low-energy explosion.

One important difference between SNe~Iax and SNe~Ia is that we have
imaged the probable progenitor system of a SN~Iax \citep[hereafter,
\citetalias{McCully14:12z}]{McCully14:12z}, while no progenitor system
of a SN~Ia has yet been directly observed.  For SN~2012Z,
\citetalias{McCully14:12z} detected a luminous, blue source in
pre-explosion {\it Hubble Space Telescope} ({\it HST}) images at the
SN position.  Their favored interpretation is that this source is the
nondegenerate He companion star to a C/O white dwarf (WD), as
originally predicted by \citetalias{Foley13:iax} as the likely
progenitor scenario for SNe~Iax.  However, it is also possible that
the light came from a massive star that exploded to cause SN~2012Z or
an accretion disk around the exploding WD.

An important prediction from both simple energetic arguments
\citep{Foley08:08ha, Foley09:08ha, Foley13:iax, McCully14:iax} or from
detailed explosion models \citep{Jordan12, Kromer13, Fink14} is that
at least some of the time, a SN~Iax, if coming from a C/O WD, should
leave behind a bound remnant star.  Since the remnant retains a
significant amount of radioactive material after the explosion
\citep{Kromer13}, this star will likely become quite luminous on the
timescale of years to decades (Bildsten et~al., in preparation).

Using {\it HST} images that include the location of SN~2008ha, a
low-luminosity SN~Iax \citep{Foley09:08ha, Foley10:08ha, Valenti09},
from 4 years after the explosion, \citet{Foley14:08ha} detected a
luminous, red source at the position of the SN.  Since these images
came after the explosion, the source could be a thermally pulsating
asymptotic giant branch (TP-AGB) companion star or a bound remnant
star.  If the sources detected at the positions of SNe~2008ha and
2012Z are both companion stars, SNe~Iax must have a diverse set of
progenitor systems.

Until recently, SN~2008ge was the only other SN~Iax with pre-explosion
{\it HST} imaging \citep{Foley10:08ge}.  These data are not
particularly deep (3$\sigma$ limit of $M_{V} > -6.9$~mag), but rule
out particularly massive stars as potential progenitors.  Notably,
SN~2008ge occurred in an S0 galaxy with no star formation to deep
limits \citep{Foley10:08ge}, which indirectly disfavors a massive star
progenitor.

In this {\it Letter}, we present deep pre- and post-explosion {\it
  HST} images of SN~2014dt, a SN~Iax discovered in M61 at a distance
of only 12.3~Mpc.  By comparing the pre- and post-explosion images, we
precisely determine the position of the SN in the pre-explosion
images.  We do not detect any star at that position with a 3$\sigma$
limit of $M_{\rm F450W} > -5.0$~mag.

%%%%%%%%%%%%%%%%%%%%
%%  Observations  %%
%%%%%%%%%%%%%%%%%%%%

\section{Observations and Data Reduction}\label{s:obs}

SN~2014dt was detected in M61 on 2014 October 29.8 (all times are UT)
at 13.6~mag by \citet{Nakano14} and promptly classified as a SN~Iax by
\citet{Ochner14} from a spectrum obtained 2014 October 31.2.  The SN
was past peak at discovery and there are no recent nondetections which
constrain the date of explosion.

M61 is in the Virgo cluster and has a distance of 12.3~Mpc ($\mu =
30.45 \pm 0.10$~mag) as determined by with the expanding photosphere
method (EPM) applied to SN~2008in, also in M61 \citep{Bose14}.  This
is consistent with the Tully-Fisher distance of 11.0~Mpc \citep[$\mu =
30.21 \pm 0.70$~mag;][]{Schoeniger97}, and the redshift-derived
distance (corrected for Virgo infall) of 13.1~Mpc ($30.59 \pm
0.16$~mag).  Here we assume the EPM distance modulus, but increase the
uncertainty to 0.24~mag, corresponding to the offset between the EPM
and Tully-Fisher measurements; this range also includes the
redshift-derived distance.

\begin{figure}
\begin{center}
\epsscale{0.9}
\rotatebox{90}{
\plotone{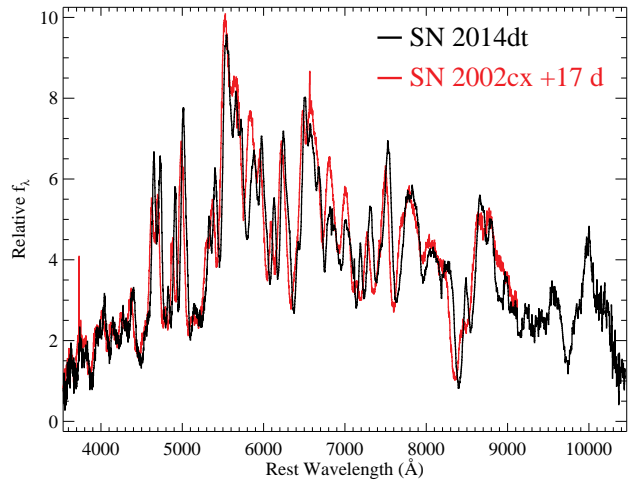}}
\caption{Optical spectrum of SN~2014dt (black) compared to that of
  SN~2002cx at +17~d (red).}\label{f:spec}
\end{center}
\end{figure}

On 18.6 November 2014, we obtained a low-resolution spectrum of
SN~2014dt with the Kast double spectrograph \citep{Miller93} on the
Shane 3~m telescope at Lick Observatory.  Standard CCD processing and
spectrum extraction were accomplished with IRAF\footnote{IRAF: the
  Image Reduction and Analysis Facility is distributed by the National
  Optical Astronomy Observatory, which is operated by the Association
  of Universities for Research in Astronomy, Inc.\ (AURA) under
  cooperative agreement with the National Science Foundation (NSF).}.
The data were extracted using the optimal algorithm of
\citet{Horne86}.  Low-order polynomial fits to calibration-lamp
spectra were used to establish the wavelength scale, and small
adjustments derived from night-sky lines in the object frames were
applied.  We employed our own IDL routines to flux calibrate the data
and remove telluric lines using the well-exposed continua of
spectrophotometric standards \citep{Wade88, Foley03,
  Silverman12:bsnip}.  The spectrum is presented in \autoref{f:spec}.

The position of SN~2014dt was imaged by {\it HST}/WFPC2 on 2001 July
27.10 (Program GO--9042; PI Smartt) in F450W (roughly $B$) and F814W
(roughly $I$), each for 460~s.  We obtained drizzled mosaics from the
Hubble Legacy Archive.

We also imaged SN~2014dt with {\it HST}/WFC3/UVIS in F438W (roughly
$B$ and well matched to the pre-explosion F450W image) on 2014
November 18.89 (Program GO--13683; PI Van Dyk).  We took 20 short
(20~s) exposures so as to not saturate the SN.

We combined exposures (including cosmic ray rejection) using
AstroDrizzle.  We drizzled the images to the native scale of WFC3,
$0{\farcs}04$ pixel$^{-1}$.  We also ran the individual flt images of
the SN through Dolphot, an extension of HSTPhot \citep{Dolphin00}, and
measured the brightness to be $m_{\rm F438W} = 16.483 \pm 0.001$ mag.

Using 30 stars in common between the F450W pre-explosion and F438W
post-explosion images, we computed a relative astrometric solution
between the two images and precisely determined the position of
SN~2014dt in the pre-explosion image.  The position of SN~2014dt has
uncertainties of 0.13 and 0.10 pixels ($0{\farcs}006$ and
$0{\farcs}005$) in the horizontal and vertical directions,
respectively.  After the initial astrometric match, we are able to
transform between the two pre-explosion images with negligible
additional positional uncertainty.

\begin{figure*}
\begin{center}
\epsscale{1.15}
\rotatebox{0}{
\plotone{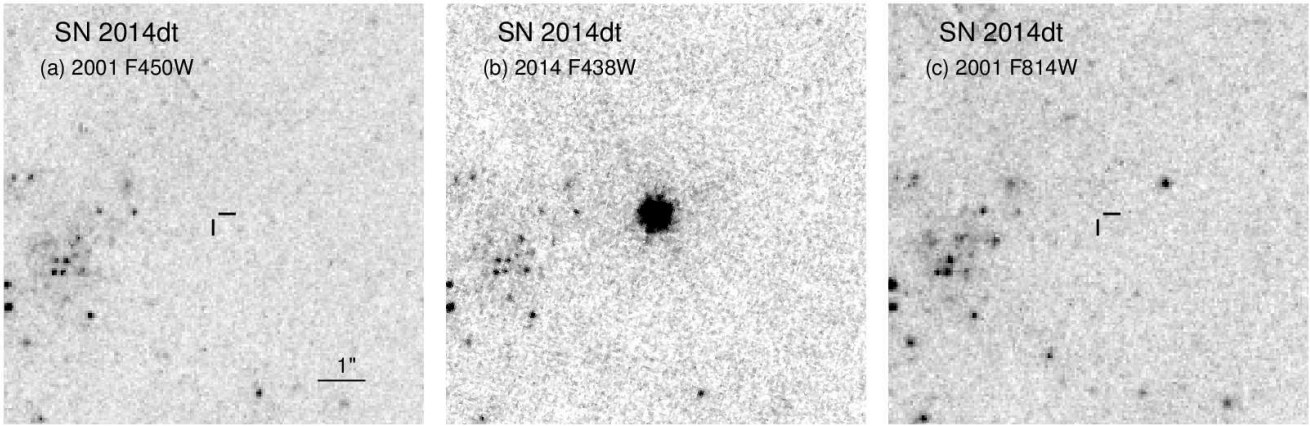}}
\caption{{\it HST}/WFC3 F438W image of SN~2014dt (b) and its
  surrounding environment.  Aligned pre-explosion {\it HST}/WFPC2
  F450W (a) and F814W (c) PC1-chip images are also shown to the same scale.
  The position of SN~2014dt in the pre-explosion images is marked.  No
  star is detected at the SN location in either
  pre-explosion image.}\label{f:finder}
\end{center}
\end{figure*}

Pre- and post-explosion images of SN~2014dt and its surrounding
environment are shown in \autoref{f:finder}.

We conducted Dolphot point-spread function photometry on the
individual pre-explosion {\it HST} flt frames, using the suggested
Dolphot parameters for WFPC2.  We only consider photometry with a
``flag'' of 0 and an ``object type'' of 1 (point sources) in our
results.

There is no source detected at the position of SN~2014dt in either
pre-explosion image.  The closest source, detected by Dolphot at
$3.6\sigma$ with 25.61~mag in F450W only, is $1{\farcs}64$ (97~pc) to
the southwest of the SN position.  The next closest, detected at
$3.6\sigma$ with 24.60~mag in F814W only, is $3{\farcs}21$ (190~pc) to
the southeast.  The closest source detected in both bands (with
24.06~mag, $11.6\sigma$ at F450W; 23.47~mag, $9.4\sigma$ at F814W) is
$3{\farcs}59$ (210~pc) to the northeast.  The 3$\sigma$ limiting
magnitude corresponds to $m_{\rm F450W} > 25.7$ and $m_{\rm F814W} >
24.8$~mag, respectively.

%%%%%%%%%%%%%%%%
%%  Analysis  %%
%%%%%%%%%%%%%%%%

\section{Analysis}\label{s:anal}

SN~2014dt is spectroscopically a SN~Iax.  The spectrum presented in
\autoref{f:spec} is compared to one of SN~2002cx, the prototypical
member of the class \citep{Filippenko03:02cx, Li03:02cx}, at 17~days
after maximum brightness.  They are nearly identical except SN~2014dt
appears to have a slightly lower ejecta velocity.  Importantly, there
is no indication of hydrogen in the spectrum.

At $D = 12.3$~Mpc, SN~2014dt had $M \approx -17.1$~mag at discovery.
Using standard SN~Iax light curves \citepalias{Foley13:iax}, SN~2014dt
likely peaked at $M \approx -18$~mag, also consistent with other
SNe~Iax.

Although it is difficult to measure host-galaxy reddening from SN~Iax
colors \citepalias{Foley13:iax}, SN~2014dt does not appear to have any
host-galaxy reddening.  The SN has a reasonably blue continuum and
there is no Na~D absorption in our high signal-to-noise ratio
low-resolution spectrum.  This is consistent with the expected
reddening from a relatively clean part of a face-on spiral galaxy.

\begin{figure}
\begin{center}
\epsscale{1.15}
\rotatebox{0}{
\plotone{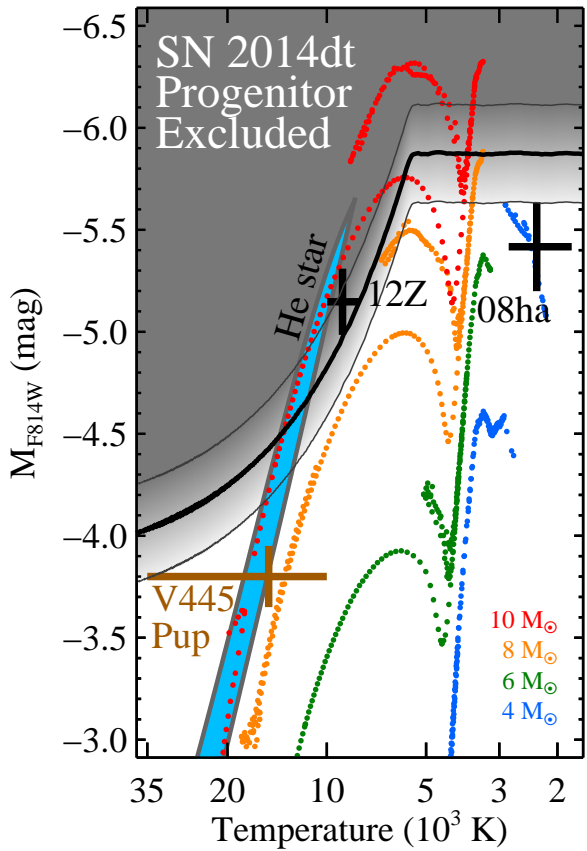}}
\caption{Hertzsprung-Russell diagram for SN~Iax progenitor systems.
  SN~2014dt is not detected, and the region of the H-R diagram
  excluded is shaded dark gray.  The allowed region is white.  The
  black curve represents the single-band 3$\sigma$ limits for black
  body sources.  The gray gradient band indicates the additional
  1$\sigma$ distance-modulus uncertainty.  The white region is allowed
  for the progenitor system.  The SN~2012Z progenitor system (black
  cross; \citetalias{McCully14:12z}) and the V445~Pup progenitor
  system (brown cross; F814W magnitude estimated from temperature and
  $V$-band magnitude; \citealt{Woudt09}) are shown.  The source
  coincident with SN~2008ha in post-explosion images is also marked
  with a black cross.  This source may be the companion star or the
  bound remnant star \citep{Foley14:08ha}.  Also plotted are stellar
  evolution tracks for stars with initial masses of 4, 6, 8, and
  10~M$_{\sun}$ \citep{Bertelli09} and the region predicted for one
  set of He-star progenitor models from \citet[blue
  region]{Liu10}.}\label{f:hr}
\end{center}
\end{figure}

No sources were detected at the position of SN~2014dt in pre-explosion
{\it HST} images.  With our given distance modulus ($\mu =
30.45$~mag), and our Milky Way and host-galaxy reddening estimates
($E(B-V) = 0.02$ and 0~mag, respectively), the 3$\sigma$ limits
correspond to $M_{\rm F450W} > -5.0$ and $M_{\rm F814} > -5.9$~mag.
These limits, converted into blackbody temperature and F814W
luminosity limits, are presented in \autoref{f:hr}.  They are similar
to the observed brightness of the probable SN~2012Z progenitor system
($M_{\rm F435W} = -5.43 \pm 0.15$ and $M_{\rm F814W} = -5.24 \pm
0.16$~mag).  Formally, the SN~2014dt limits are 0.40~mag deeper than
the SN~2012Z progenitor system detection in F435W/F450W.  However,
this is only 1.4$\sigma$ different when including uncertainties in the
SN~2012Z progenitor system photometry and the distance to M61.  We
therefore cannot rule out a progenitor system for SN~2014dt that is
similar to that of SN~2012Z.

Our limits are not quite deep enough to rule out a progenitor system
similar to that of the source coincident with SN~2008ha at late times
($M_{\rm F814W} = -5.42 \pm 0.22$~mag).  Moreover, a system similar to
that of V445~Pup (the only known He nova; \citealt{Kato03}), thought
to be a C/O WD with a He-star companion \citep{Kato08,Woudt09}, would
be barely undetected in the pre-explosion images.  Therefore,
SN~2014dt could have a progenitor system similar to any of these
comparison objects.

Nonetheless, a large number of potential progenitor systems have been
excluded.  Using a projected offset of 2.3~kpc, and a metallicity map
of M61 \citep{Pilyugin14}, we find a metallicity at the SN position of
$12 + \log ({\rm O/H}) = 8.68$, which on the scale adopted by
\citet{Pilyugin14} is roughly 1.5 times solar.  From single-star
evolutionary models at this metallicity \citep{Bertelli09}, the
pre-explosion limits are inconsistent with systems containing a red
giant (RG) or horizontal branch (HB) star with $M_{\rm init} \gtrsim
8~ {\rm M}_{\sun}$ or main-sequence stars with $M_{\rm init} \gtrsim
16~ {\rm M}_{\sun}$.  Of course, these kinds of stars are unlikely to
have been the progenitor because of the lack of hydrogen in the SN
spectrum; nonetheless, these limits apply to companion stars as well.

The data do not exclude very hot Wolf-Rayet stars, which can be
relatively faint at optical wavelengths \citep{Shara13}.  However,
Wolf-Rayet stars would likely be physically close to other massive
stars.  SN~2014dt exploded in a region that is \about 100~pc from any
detected sources and $\gtrsim$200~pc from any particularly bright
sources, making the Wolf-Rayet scenario less likely.

%%%%%%%%%%%%%%%%%%%%%%%%%%%%%%%%
%%  Discussion & Conclusions  %%
%%%%%%%%%%%%%%%%%%%%%%%%%%%%%%%%

\section{Discussion and Conclusions}\label{s:disc}

Using {\it HST} images obtained 13 years before SN~2014dt, a clear
SN~Iax, we place limits on its progenitor system.  This is the third
SN~Iax with pre-explosion {\it HST} images and the second deepest
(after SN~2012Z).

With these data, we are able to exclude many massive stars as
progenitors for SN~2014dt.  Specifically, main-sequence stars with
$M_{\rm init} \gtrsim 16~{\rm M}_{\sun}$ and RG/HB stars with $M_{\rm
  init} \gtrsim 8~{\rm M}_{\sun}$ are excluded.  Some Wolf-Rayet stars
are still allowed, but the lack of bright nearby stars make a
Wolf-Rayet progenitor unlikely.

If the progenitor system of SN~2012Z were in M61, we would perhaps
marginally detect it.  Since we did not detect any sources at the
position of SN~2014dt, its progenitor system was likely fainter (in
$B$) than that of SN~2012Z.

Many of the He stars in the He-star--C/O-WD progenitor models of
\citet{Liu10} have roughly the same luminosity but varying
temperature.  Since the He-star spectral energy distribution peaks in
the ultraviolet, the effective temperature dictates the brightness in
the {\it HST} bands.  A large region of the He-star parameter space is
still allowed by the current data.  However, future data, particularly
at shorter UV wavelengths, should be able to place strong constraints
on the existence of such a star.

The pre-explosion data also do not exclude a red giant star similar to
what may be the companion of SN~2008ha, although the data are within
\about 0.5~mag of being able to detect such a star.  Again, future
data should be able to easily detect a similar star, although as to
the case of SN~2008ha, the interpretation of such a detection would be
somewhat ambiguous.

There are now three SNe~Iax with reasonably deep pre-explosion images.
A probable luminous blue progenitor system was detected for SN~2012Z
\citepalias{McCully14:12z}, the SN with the deepest pre-explosion
data, while no progenitor was detected for either SN~2008ge
\citep{Foley10:08ge} or 2014dt.  \citet{Foley14:08ha} detected a
luminous red source coincident with SN~2008ha 4 years after the
explosion that could be a companion star or the bound remnant of the
progenitor WD.  These four SNe independently and jointly rule out
massive stars as SN~Iax progenitors.

SN~2014dt, being at $D = 12.3$~Mpc, is likely the closest SN~Iax yet
discovered (SN~2010ae was at 12.9~Mpc; \citealt{Stritzinger14:10ae}),
making it a great candidate for long-term monitoring.  Of the 5
SNe~Iax discovered within \about 20~Mpc (SNe~2008ge, 2008ha, 2010ae,
2010el, and 2014dt), SN~2014dt is the best SN for such observations;
SNe~2010ae and 2010el have significant reddening, SN~2008ge was close
to a bright galactic nucleus, and SN~2008ha was the faintest and most
distant.  SNe~Iax fade by \about 12~mag in the first two years
\citep{McCully14:iax}.  SN~2014dt peaked at $M \approx -18$~mag, and
will be roughly as bright as the pre-explosion image limits in two
years.  Around that time, deep images may be able to detect emission
from either a companion star or the WD remnant.

\begin{acknowledgments} 

  {\it Facility:} \facility{Hubble Space Telescope (WFPC2, WFC3)},
  \facility{Lick Shane (Kast)}

\bigskip

Based on observations made with the NASA/ESA {\it Hubble Space
  Telescope}, obtained at the Space Telescope Science Institute
(STScI), which is operated by the Association of Universities for
Research in Astronomy, Inc., under NASA contract NAS 5--26555.  These
observations are associated with and funded through program GO-13683.
Data were obtained through the Hubble Legacy Archive, which is a
collaboration between the Space Telescope Science Institute
(STScI/NASA), the Space Telescope European Coordinating Facility
(ST-ECF/ESA) and the Canadian Astronomy Data Centre (CADC/NRC/CSA).

SN~Iax research at the University of Illinois is supported in part
through NASA/{\it HST} grant GO--12999.01.  This research at Rutgers
University was supported through NASA/{\it HST} grant GO--12913.01 and
National Science Foundation (NSF) CAREER award AST--0847157 to S.W.J.
A.A.M. acknowledges support for this work by NASA from Hubble
Fellowship grant HST--HF--51325.01, awarded by STScI.  A.V.F.'s group
at U.C. Berkeley is supported by the Christopher R.\ Redlich Fund, the
TABASGO Foundation, and NSF grant AST--1211916.

\end{acknowledgments}

\bibliographystyle{fapj}
\bibliography{../astro_refs}

%\eject

\end{document}